\documentstyle[aps,multicol,psfig,epsf,epsfig]{revtex}
\begin{document}
\draft
\tightenlines

\title{Numerical study of persistence in models 
with absorbing states}
\author{
Ezequiel V. Albano $^1$ and Miguel A. Mu\~noz $^2$}
\address{
$^1$ Instituto de Investigaciones Fisicoqu\'{\i}micas
Te\'oricas y Aplicadas, (INIFTA), CONICET
, UNLP, CIC (Bs.As.).
Sucursal 4, Casilla de Correo 16, (1900)
La Plata, ARGENTINA. \\        
$^2$ Institute {\em Carlos I} for Theoretical and Computational Physics\\
and Departamento de E. y F{\'\i}sica de la Materia,\\
Universidad de Granada
 18071 Granada, Spain.\\
}
\date{\today}

\maketitle

\begin{abstract}
\indent Extensive Monte Carlo simulations are performed 
in order to evaluate both the local ($\theta_{l}$) and global 
($\theta_{g}$) persistence exponents in the Ziff-Gulari-Barshad
(ZGB) ( Phy. Rev. Lett. {\bf 56}, 2553, (1986) ) 
irreversible reaction model. In two dimensions and
for the second-order irreversible phase transition (IPT) we find, 
that both the local and the global persistence exhibit 
power-law behavior with a crossover between two different time regimes. 
On the other hand, at the first-order IPT, characteristic of the
ZGB, active sites are short lived and the persistence decays
more abruptly, not being clear whether 
it shows power law behavior or not.      
In order to analyze universality issues,
 we have also studied another model with absorbing states, the contact
process, and evaluated the local persistence exponent in dimensions
from $1$ to $4$. A striking apparent {\it super-universality} is reported:
the local persistence exponent seems to coincide in  both
one and two dimensional systems. 
\end{abstract}

\begin{multicols}{2}

\section{Introduction}

  An avalanche of articles devoted to the study of persistence
in different physical contexts has appeared recently in the literature.
  From a theoretical point of view, the appearance of 
new nontrivial critical exponents as: i) the 
``short-time'' exponent, $\lambda$, needed 
to characterize the two-time correlation functions in systems relaxing
in the process of quenching from infinitely 
high temperatures to the critical temperature $T_{c}$
 \cite{jans,huse}, and ii) the  
global persistence exponent $\theta_{g}$, related to the probability 
$P_{g}(t) \propto t^{-\theta_{g}}$, that the global order
parameter, (e.g. the magnetization in a ferromagnet), has not changed 
sign up to a certain time $t$ after a quench to $T_c$
\cite{maju}, constituted an exciting surprise.
 A number of theoretical,
computational, and even experimental analysis have been performed 
afterwards. 
 From the theoretical viewpoint 
it has been shown that for processes where the global order 
parameter dynamics is Markovian, that the global persistence
exponent can be related to other ``traditional'' critical
exponents: in particular, the following scaling law holds \cite{maju}:
\begin{equation}
\theta_{g}  z = \lambda -  d + 1 - \frac{\eta}{2},
\end{equation}
 where $d$ is the dimensionality and $\eta$ is the static 
critical exponent of the order parameter correlation function. 
Nevertheless, the required hypothesis is not typically satisfied
 for most of the usually studied statistical models; i.e. the
dynamics of the global order parameter can be argued to be
generically non Markovian. In these more generic cases
Eq. (1) does not hold and 
  $\theta_{g}$ is a new independent non-trivial exponent
\cite{maju}. 
In general, persistence exponents depend on the system evolution
as a whole and, therefore, analytical estimations of them are scarce and
difficult \cite{derr,Derrida,balda}.
 One important theoretical contribution is  by
Majumdar and Sire, who proposed a method to calculate autocorrelation
functions perturbatively for non-Markovian processes 
(perturbing around a Gaussian and Markovian  process \cite{SMR,MS}).

Recently, other  similar non-trivial
 exponents have been undercovered and measured in different
 systems. Some of them are:
\begin{itemize}
\item
 (i) The local persistence probability 
$P_{l}\propto t^{- \theta_{l}}$ defined as the probability 
that the local order parameter at a given point $\vec{x}$ has never 
changed sign after the initial time (e.g. the probability than one
spin has never been flipped in a ferromagnet) \cite{maju,stau}.
Local persistence exponents have been measured in real
systems as liquid crystals or soap bubbles (see \cite{SMR} and
references therein). 

\item (ii) The block persistence probability 
$P_{L}\propto t^{-\theta_{L}}$ defined as
the probability that the order parameter integrated over a block 
of linear size $L$ has never changed sign
since the initial time in a phase-ordering process at finite 
temperature \cite{cuei,sire};

\item (iii) The generalized 
persistence probabilities and exponents
introduced by Dornic and Godr\'eche \cite{DG,balda}.

\item (iv) The persistence exponents for domains. Found
first by Krapivsky and Ben-Naim in Ising system and subsequently in
diffusion by Majumdar and Cornell \cite{iv}. These can also be
 generalized to
other different types of pattern:
each pattern decays with a 
different exponent, a single spin or the global spin being just two
specific patterns. 

\item (v) The persistence with partial
survival first studied by Majumdar and Bray and subsequently by
other authors \cite{v}. Recently it has generated a lot of interests in the
context of inelastic collapse in granular materials. 

\end{itemize}
  For a review of persistence
studies in reversible systems see e.g. \cite{sire,DG,CSI}.

Even though considerable effort has been devoted to
 the study of persistence in various models exhibiting 
reversible phase transitions
(see  
\cite{jans,huse,maju,derr,stau,cuei,sire,mnyh,stauffer,schu,newm,hinr} 
and references therein) as well as in
diffusion-reaction systems
 (see \cite{card,majum} and references therein), 
far less numerous are the studies of persistence in systems 
exhibiting irreversible phase transitions (IPT's) and, in particular, 
in systems presenting a critical point separating an active 
from an absorbing phase (some exceptions can be found
in  \cite{hinrin,oerd,howard}).  In this context,
recently, Hinrichsen and Kodusely \cite{hinrin}
have performed a numerical study of the local and global persistence 
in $(1 + 1)-$dimensional directed percolation (DP) \cite{reviews}
(see also \cite{Hinrichsen}). 
They found that the local persistence probability at the critical point
(separating the absorbing from the active phase)
 decays as a power law with an exponent
 ${\theta_{l}}^{DP}\approx 1.5$. 
Also, global persistence measurements seemed to be consistent
with ${\theta_{g}}^{DP} \approx 1.5$,
a result that has to be taken with some caution since, on
the one hand, the evaluation of ${\theta_{g}}^{DP}$ requires 
extensive simulations (see the criticisms to this result
in \cite{oerd}) and on the other hand, in all the
 previously known models 
(mainly for equilibrium critical systems)  
${\theta_{g}} < {\theta}_{l}$ \cite{maju}. Furthermore, it
was conjectured that both $\theta_{g}$ and $\theta _{l}$ are 
indeed universal exponents characterizing the DP universality class.
Another interesting point, is that the persistence exponents in this
case seem to be independent on the initial condition at odds with
what happens in other well known cases as, for example, 
the two-dimensional Ising model with Glauber dynamics \cite{Derrida}.
  Finally, Hinrichsen and Koduvely 
succeeded  in relating persistence exponents in directed percolation
to certain ``return'' probabilities with an absorbing boundary 
or an active source \cite{hinr}. 
  Another  important contribution to the study of this IPT was
 presented in the recent paper by Oerding and van Wijland, in which the 
global persistence exponent has been calculated  analytically 
by combining the perturbative method developed by 
Majumdar and Sire \cite{MS} with standard renormalization group
 techniques \cite{oerd}. 
These same authors have thrown some doubts on the validity of the 
numerical estimations obtained by Hinrichsen and Koduvely (as will
be discussed in section III).

In order to contribute further to the understanding of persistence
in systems with irreversible phase transitions and try
to shed some light on the aforementioned issues, 
the present manuscript is devoted to the numerical study and evaluation 
of persistence exponents of the Ziff-Gulari-Barshad (ZGB) model
for the catalytic oxidation of $CO$ \cite{ziff}, as well as  
in the contact process \cite{CP,reviews} in various dimensions. 
The ZGB model in $d = 2-$dimensions has a twofold advantage: 
(i) it exhibits  a second-order IPT unambiguously placed at the 
DP universality class \cite{jens}, and (ii) it also exhibits a 
first-order IPT \cite{ziff} where, as expected,
the system does not show scaling neither
 universal behavior \cite{monetti}. 
Therefore, the ZGB model provides a suitable framework
for the study of the persistence in both,
 first and second order irreversible critical points.
 Furthermore, the present study extends the investigation 
of persistence exponents to higher dimensions in IPT's,
by performing numerical studies of the contact process.

The manuscript is organized as follows: 
In section II the ZGB model and the simulation method are
described in detail. In section III we define the local and 
global persistence, while section IV
is devoted to the presentation and discussion
 of the main results for both the ZGB model
 and the contact process. 
Finally in section V the main conclusions are presented.

\section{The ZGB model and the simulation method}

The ZGB model mimics the catalytic oxidation of carbon monoxide 
on a transition metal surface \cite{ziff}, namely
$2CO + O_{2} \rightarrow 2CO_{2}$.
The lattice-gas version of the ZGB model is also known as the 
monomer-dimer model (MD), where $A \equiv CO$ is the monomer because 
it needs a single adsorption site on the surface, and 
$B_{2} \equiv O_{2}$ is the dimer which adsorbs dissociatively, 
and consequently requires two neighboring sites on the catalyst
surface to be adsorbed.
 It is assumed that the monomer-dimer reaction 
proceeds according to the Langmuir-Hinshelwood mechanism:

\begin{equation}
A(g) + S \rightarrow A(a),  
\end{equation}

\begin{equation}
B_2(g) + 2S \rightarrow 2B(a),
\end{equation}

\begin{equation}
A(a) + B(a) \rightarrow AB(g) + 2S, 
\end{equation}
where $S$ is an empty site on the surface, while $(a)$ and $(g)$
refer to the adsorbed and gas phase, respectively.

 The MD model uses a square lattice to represent the
catalytic surface. The Monte Carlo algorithm for its
simulation is as follows : i) an $A$ or $B_2$ molecule
is selected randomly with relative probabilities $Y_A$ and $Y_B$,
respectively.  These probabilities are the relative impingement
rates of both species, which are proportional to their partial
pressures. Due to the normalization, $Y_A$ + $Y_B$ = 1, the model
is characterized by a single parameter, say $Y_A$.
  If the selected species is $A$, a surface site 
is selected at random, and if that site is  vacant,  $A$ is
adsorbed on it [Eq.(2)].
 Otherwise, if that site is occupied, the trial ends
and a new molecule is selected. If the selected species is $B_2$,
a pair of nearest-neighbor sites is randomly chosen and the molecule is
adsorbed on them only if they are both vacant [Eq.(3)]. ii) After each
adsorption event, the nearest neighbors 
of the added molecule are examined
in order to account for the reaction given by Eq.(4).
 If more than one pair
$[B(a),A(a)]$ are identified, one of them is randomly selected and
removed from the surface (for more details on the MD and
the simulation technique see, e.g. \cite{ziff,hcr}).

 Interest in the MD model arises due to its
rich and complex phenomenology. In fact,
in two dimensions and for the asymptotic regime
$(t \rightarrow \infty)$, the system reaches a stationary state
whose nature depends solely on the parameter $Y_A$.
In fact, for $Y_A \leq Y_{1A} \cong 0.387368$
($Y_A$ $\geq$ $Y_{2A}$ $\cong$ 0.52554)
the surface becomes irreversibly poisoned by $B$ $(A)$ particles.
 On the other hand, for $Y_{1A} < Y_A < Y_{2A}$ a steady 
state with sustained production of $AB$ is generated.
Figure 1 shows plots of the rate of $AB$ production ($R_{AB})$ and the
surface coverages of $A$ $(\theta_A)$ and $B$ $(\theta_B)$ 
versus $Y_A$, respectively.
Just at $Y_{1A}$ and $Y_{2A}$ the MD model exhibits
irreversible phase transitions (IPT's)
between the reactive regime and poisoned states, which 
are of second and first order respectively. 
The second-order IPT  belongs to the universality class of
directed percolation \cite{reviews}
 and is rather well understood \cite{jens}.
Furthermore, as it is shown in Figure 1, when $Y_A$ increases 
towards $Y_{2A}$ the catalytic activity increases, 
but when  $Y_{2A}$ is reached large $A$ clusters
suddenly emerge and cover the whole lattice. The transition
occurs abruptly, with discontinuities of the coverages and activity,
unveiling its first-order nature.

\section{Local and global persistence}

In a system, such as directed percolation,
 possessing absorbing states,
the {\it local persistence } probability 
can be measured by starting the system  
with a homogeneous random initial conditions,
and evaluating the probability that a given, originally inactive,
site has not become active up to time $t$ \cite{hinr}. 
It should be noticed that this quantity depends upon the whole
system evolution up to time $t$, and therefore it can be 
enviewed as an infinite-point correlation
function in the time direction. Consequently
${P_{l}}(t)$ is a rather non-trivial quantity,
difficult to study analytically as discussed before.
Numerical simulations in a 
$(1 + 1)$-dimensional DP model shows that ${P_{l}}(t)$ 
decays algebraically at criticality: 
\begin{equation}
{P_{l}(t) \propto t^{-\theta_{l}}}
\end{equation}
with  ${\theta_{l}}\approx 1.5$ \cite{hinr}.
In the case of the ZGB model, and for the continuous phase transition,
the absorbing state corresponds to the
surface of the  catalyst fully covered by $B$-species. Such species
plays the role of inactive sites (for the other transition, i.e. the
first-order one, the roles of the $A$ and $B$ particles are
exchanged).  We define 
$P_{l}^{ZGB}(t)$ as the probability that a given inactive
site does not become active up to time $t$. 
Consequently, for the second-order (first order) transition 
 simulations are started with lattices 
partially covered at random by $A$-species ($B$-species) 
 ($\theta_{A(B)}(t=0)$).
The only mechanism capable to activating a site is the reaction 
$A + B \rightarrow AB$ as specified by equation 4.

We have numerically evaluated the persistence 
distribution function $D_{l}^{ZGB}(t)$, that is, the
probability for a given site to become active in the time interval 
between $t$ and $t+dt$. Accordingly, one has 
\begin{equation}
D_{l}^{ZGB}( t ) \propto  t^{-(\theta_{l}^{ZGB} + 1)},  
\end{equation}
where $\theta_{l}^{ZGB}$ is the persistence exponent
 (${D_{l}}^{ZGB}(t)$ is simply the time derivative 
of $P_{l}^{ZGB}$).

In systems exhibiting reversible phase transitions, 
the {\it global persistence} is usually defined as the
probability that the global order parameter 
(e.g. the total magnetization in the Ising model)
does not change sign up to time $t$. In contrast, 
for systems with absorbing states, such as DP,
the global order parameter given by the density of 
active sites $\rho_{AS}(t)$, is a strictly positive
quantity and therefore the standard definition is 
not applicable. Hinrichsen and Koduvely \cite{hinr} have
proposed that, instead, one can evaluate the probability 
that the deviation of the order parameter from its mean value, 
$\Delta\rho_{AS}(t) = \rho_{AS}(t) - <\rho_{AS}(t)>$, 
does not change sign up to time $t$. 
 In the present study, and close to the second-order IPT, 
we have considered       
\begin{equation}
\rho_{AS}(t) = 1 - {\theta_{B}} = {\theta_{A} + \theta_{V}}  
\end{equation}
where ${\theta_{V}}$, $\theta_{A}$ and $\theta_{B}$ are the 
density of empty and occupied by $A$ and $B-$species sites,
 respectively. 
However, in finite systems as directed percolation the 
deviation probability 
depends on the sign of $\Delta \rho_{AS}(t)$. This asymmetry
should vanish in the thermodynamic limit where 
$\Delta\rho_{AS}(t)$ becomes a Gaussian process. 
In finite systems, however, the renormalized variance of the 
fluctuations is no longer constant but increases with the
value of $\Delta\rho_{AS}(t)$ causing the effective time scales 
for positive and negative fluctuations to be different 
\cite{hinr,oerd}. Therefore, $P_{g}^{(-)}(t)$ 
($P_{g}^{(+)}(t)$) is 
defined as the probability of $\Delta\rho_{AS}(t)$
to remain negative (positive) from an initial time $t_{in}$
up to time $t$. We 
have also evaluated the distribution function $D_g^{ZGB}(t)$, 
i.e., the probability that $\Delta \rho_{AS}(t)$ changes sign
in the time interval between $t$ and $t+dt$.
 In order to evaluate  $\Delta\rho_{AS}(t)$ we have
first to calculate a ``calibration curve'' given by
$\rho_{AS} \propto t^{-\theta}$ where $\theta\simeq 0.4505$ is
the corresponding DP exponent in
 $(2 + 1)-$dimensions \cite{brief,reviews},
describing the time decay of a homogeneous initial condition
at criticality: every-time $\rho_{AS}(t)$ intersects this
curve, $\Delta\rho_{AS}(t)$ changes sign.

 A second important issue related to the global persistence 
exponent, worth to discuss before proceeding, is 
the time regime in which it should be measured. 
Following \cite{oerd}, the global exponent is well defined 
in the regime in which the 
dependence on the initial state has been erased and, on the other 
hand, there is an upper bound to the validity of measurements, given 
by the upper cutoff induced by the finite system size. 
Putting together these two constraints \cite{oerd}
\begin{equation}
L^z \gg t \gg \theta_{B}^{2z/(d-\eta)}
\end{equation}
(for general systems with absorbing states, 
one has to substitute $\theta_B$ by the initial
density of active sites).

Using the DP values for the exponents, this implies 
 $18300 \gg t \gg  1$ for system size $L=256$,
and   $65000 \gg t \gg  1$ for $L = 512$.   
Therefore, as a safe, conservative limit, we start measuring at
  $t_{in} = 100$.  

It is worth stressing that in order to obtain reliable data
for persistence exponents one has to perform extensive
 simulations.
The evaluation of the local persistence requires long ranges
(up to $10^4$ Monte Carlo steps),
 and reliable statistic is obtained averaging over
$10^4$ ($2.5 \times 10^3$) different initial configurations for 
$L = 128$ ($L = 256$). On the other hand, each
single point for the global persistence distribution requires
a whole run; therefore good statistics is much tougher to obtain that
for local persistence. 
We averaged over $10^6$ and $2.10 \times^5$ different
runs for $L = 256$ and $L = 512$, respectively.

\section{Results and discussion.}
 
In this section we report our main numerical findings.
The value of the second order
critical point of the ZGB model
 reported previously,
 $Y_A=0.387368$ \cite{jens}, corresponds to the 
infinite size limit; while for the system sizes
 we will consider, namely $128*128$
and $256*256$, finite size corrections shift the critical value
to slightly larger values: in particular, all the simulations
at the critical point reported in what
 follows are performed at $Y_A=0.3907$ \cite{jens}.

Figure 2 shows a log-log plot of $D_{l}(t)$ 
versus $t$ at the second-order IPT.
Two regimes can clearly we observed: i) The short time one,
for $t < 200$, where $D_{l}(t)$ exhibits a power law 
behavior $D_{l}(t)\propto t^{-(\theta_{l}+1)}$,
with $\theta_{l} \simeq 1.00 \pm 0.05$, and 
(ii) the asymptotic regime, for $t > 200$ , with the exponent
$\theta_{l}^{ZGB} \simeq 1.50 \pm 0.10$.   
Since the phase transition belongs in the
 DP universality class \cite{jens}, 
it is quite surprising that the obtained exponent, within error
bars, is almost the same than the one
 reported by Hinrichsen et al. 
\cite{hinr}, for a DP process in $(1 + 1)-$dimensions,
namely $\theta_{l}^{DP} \simeq 1.50 \pm 0.03$. 
This finding suggests that for DP, the difference 
between the persistence exponents in $(1 + 1)-$ and 
$(2 + 1)-$dimensions is very small.
 In addition, we have also verified the lack of dependence of these
exponents upon modifications of the initial condition,
as shown in figure 3. 
In fact, runs performed using $0.05 \leq \theta_{B}(t = 0) \leq 0.20$
show clearly  that neither the short-time regime nor the 
persistent regime depend on the initial density of 
inactive sites. This finding is in accordance
with Hinrichsen et al. observation \cite{hinr}.

In order to gain some insight on the origin of the crossover 
observed in figures 2 and 3, we have evaluated the time
dependence of the fraction of persistent sites with $NN=0$, $1$,
$2$, $3$ and $4$ empty nearest-neighbor respectively, as a function
of time, 
as it is shown in figure 4. 
It is clear that different $NN$ values 
vanish at distinct characteristic times, 
e.g. $t \simeq 10, (NN = 0)$; 
$t \simeq 15,  (NN = 1)$; $t \simeq 30, (NN = 2)$; 
$t \simeq 120, (NN = 3)$; and 
$t \simeq 300, (NN = 4 )$, respectively.
Therefore comparing the results of figures 2-3 and 4 we 
conclude that the crossover in figures 2-3 takes place when almost
all persistent sites are blocked by inactive (absorbing) ones.
The physical interpretation of the observed crossover is therefore
the following: There is an initial regime in which the persistence
of a given site is dominated by its open neighboring sites: 
particles can land directly in the neighboring sites causing 
desorption, and consequently the death of persistent sites. However,
as the number of such open sites is reduced on average to a value
close to zero, there is a crossover to the true asymptotic 
regime in which one has to wait 
for local rearrangements in order to reach persistent sites.

In order to further explore the dependence of the persistence exponents 
on the dimensionality we have also performed extensive simulations 
\cite{mamu}, of another system with absorbing states belonging
to the DP universality class, namely the contact process (CP)
 \cite{reviews,CP}. 
In this model, proceeding analogously, we observe no  crossover
analogous to the one described for the ZGB, asymptotic 
regimes can be obtained with less computational effort,
 and the  exponent values are far more accurate. We simulate 
the CP in dimensions from $1$ to $4$. The largest
 system sizes considered are $10^4$ in $d=1$, $256*256$ in $d=2$,
$50^3$ in $d=3$, and $32^4$ in $d=4$.
Our results are 
 $\theta _{l}\simeq 1.50 \pm 0.01$ for $(1 + 1)-$,  and 
$\theta_{l}\simeq 1.50 \pm 0.01$ for 
$(2 + 1)-$ dimensions \cite{mamu} (see Fig. 5 ).
 In fact, the two curves in Figure 5 are strikingly parallel.
 This fact confirms our rather strange
finding, that the local persistence exponent is (almost?) the 
same in one and two-dimensional systems with absorbing states.
It could be the case, that there is some small difference
between the exponent values in $d=1$ and $d=2$, but within
our numerical accuracy they are absolutely indistinguishable.  
 The apparent coincidence between the
results in one and two dimensions is quite intriguing, and so far, we
do not have any satisfactory explanation for this surprising result.

For the sake of completeness,
and in order to test whether this apparent {\it ``super-universality''}
 extends also to higher dimensions,  we
have also measured the local persistence exponents in $d=3$ and $d=4$
as said before.
The results for  $\theta_l$ are $1.33 \pm 0.03$ in $d=3$, 
and $1.15 \pm 0.05$
in $d=4$, indicate a monotonous diminution of the local persistence
as a function of dimensionality \cite{mamu}.

It is well known that for equilibrium models the persistence
exhibits generically a power-law behavior below the 
critical temperature \cite{sire,DG}. In order to
establish analogies with irreversible systems having absorbing
states, we have also measured the local persistence
away from criticality, i.e. within the reactive
regime of the ZGB model, as it is shown in figure 6. For
$Y_{A} = 0.40$ (i.e. very close to the critical threshold)
the power-law characteristic of the short-time regime
is still observed, but the persistent regime
shows a clear departure from the behavior observed
at criticality. Going deeper into the reactive regime,
e.g. for $Y_{A} = 0.45$, the plot shows a marked
curvature and the existence of a power-law behavior can
be ruled out. On the other hand, in the absorbing phase
persistence curves tend to a constant for large values of $t$,
 i.e. there is  a non-vanishing asymptotic  probability 
to persist indefinitely. 
This is a direct consequence of the fact that eventually the system
reaches an absorbing configuration and the dynamics is arrested. 
 Therefore, we conclude that unlike reversible
systems where a power-law behavior holds for a wide range
below the critical temperature, in systems with 
absorbing states
the power-law behavior of the persistence
is restricted to criticality.    

In what respect the global persistence, 
Figure 7 shows a double logarithmic plot of $D_{g+}^{ZGB}(t)$ 
versus $t$ obtained at the second-order IPT
for two different lattice sizes. As in the case of the local 
persistence, two time regimes observed: i) a short time regime  
($t \leq 300$) where a power law behavior with a slope 
$\simeq 1.25 \pm 0.05$, and ii) an asymptotic regime 
($t > 300$) with a power law behavior with exponent 
${\theta_{g}^{ZGB}} \simeq 2.5 \pm 0.5$.
The onset of the long time regime depends on 
$L$ as it is shows in figure 6 for $L = 256$ 
and $L = 512$, however $\theta_{g}^{ZGB}$ appears to be 
size independent.
As it was already discussed for the case of the local persistence, 
the onset of the crossover between the  different
time regimes can be linked with the results shown in figure 4. 
Consequently the asymptotic regime in
the ZGB model starts when persistent sites are almost 
surrounded by inactive sites.

Our result for the persistence exponents, 
namely $\theta_{l} < \theta_{g}$, is quite surprising
at first sight, 
since from all previously accumulated experience, 
 mainly for equilibrium 
critical systems, it is known that  
$\theta_{l} > \theta_{g}$, \cite{maju}.
It seems that irreversible critical systems 
may depart from this behavior.
 
Assuming the dynamics of the order parameter
to be Markovian, and using the known 
scaling relations, and values of the
 directed percolation universality class,
one obtains that Eq. (1) can be written as \cite{oerd,brief}
$\theta_g=1+ d/(2 z) \approx 1.565$ (in $d=2$). Our numerical
value is far from this result, implying that the process
is strongly non-Markovian.
Oerding and van Wijland have shown \cite{oerd},
 using field theoretical tools
combined with the perturbative expansion method by
Majumdar and Sire \cite{maju},
that up to first order in epsilon expansion,
$\theta_g = 2 + 0.059 \epsilon
 +\Theta(\epsilon^2)$. This implies $\theta > 2$ (at least
up to first order) as, in fact, 
is the case in our measurements.

The physical interpretation for the fact 
that $\theta_{l} < \theta_{g}$
is qualitatively the following:
at the critical point, in the activity density
 decay process it is more likely to fluctuate around the
global average density
(which corresponds to annihilating more (or less)
 activity than the average at that time),
 than invading absorbing regions in which persistent sites exist. This
is very reasonable since in general, at the critical point
and for large enough times, activity is restricted to some
patches, which become smaller and smaller as time runs. 
It is therefore, more likely to have fluctuations in the density
of these patches (controlled by the global persistence exponent)
 that to invade persistent sites (controlled by the local
exponent)
typically  surrounded by absorbing regions at large times.

The insert of figure 7 shows a log-log of 
$D_{g-}^{ZGB}(t)$. 
Here the events are short-lived making difficult the
evaluation of the persistence exponent. However, considering
larger error bars in this case, the exponent evaluated 
for $D_{g-}^{ZGB}(t)$ is consistent with the exponent 
previously calculated for $D_{g+}^{ZGB}(t)$. So, 
within the discussed limitations for the evaluation of
the exponents, we expect that both $D_{g-}^{ZGB}(t)$  
and $D_{g+}^{ZGB}(t)$ exhibit power-law behavior with the 
same exponent in agreement with the predictions
 of Oerding et al. \cite{oerd}.

Since the ZGB model also exhibits a first-order IPT 
at $Y_{2A} \simeq 0.52554$, we have evaluated the 
persistence at the coexistence point.
 Figure 8 shows a log-log plot  
$D_{l}^{ZGB}(t)$ versus $t$ as obtained for 
lattices of size $L = 512$. Since the results
are almost independent of $L$, finite size effects may be
almost negligible. 
Figure 8 also shows that at coexistence the persistent sites
are short-lived and consequently the persistence drops off 
abruptly. The data can be  fitted by a power
law with an exponent $\theta_{l} \simeq 3.0 \pm 0.3$. 
However, since the range of the fit is narrow
 and there seems to
be a systematic curvature, we cannot disregard an 
exponential (or an stretched exponential) decay for longer times. 
Further clarification of this issue will have to wait for
computationally expensive large-scale numerical simulations.
 It should be noticed that the operation of
 power-law (scale-invariance) in the dynamical critical
 properties of the first-order IPT of the ZGB model has 
recently been ruled out \cite{monetti}. But given the lack of a
general theoretical framework, we do not know whether
or not algebraic
decay of persistence functions should be expected at 
non-equilibrium first order transitions. 

\section{Conclusions}

 A numerical study of persistence in the ZGB  
dimer-monomer model for the catalyzed reaction
$A+ \frac{1}{2} B_{2} \rightarrow AB_{2}$ is presented. 
It is found that for the second-order IPT the time dependence 
of both the local and the global persistence exhibit 
a crossover between a short- and a long-time regimes. 
A physical explanation for such a crossover has been 
provided.
Persistence exponents are evaluated within the long-time 
regime giving $\theta_{l}^{ZGB} \simeq 1.50 \pm 0.01$
for the local one, and
$\theta_{g}^{ZGB} \simeq 2.5 \pm 0.5$ for the global one.
The former, which should correspond to the DP universality class 
in $(2 + 1)-$dimensions is very close to the value
reported by Hinrichsen et al. \cite{hinrin} 
for DP in $(1 + 1)-$dimensions, namely 
$\theta_{l}^{DP} \simeq 1.50 \pm 0.03$. 
 Simulations in the contact process (another model 
belonging in the same universality class) confirm this
indistinguishability between the one and two-dimensional
local persistence exponents with higher accuracy
 (see the striking parallelism of the curves in figure 5).
 This finding implies a 
theoretical and numerical puzzle: Is $\theta_{l}^{DP}$ 
 independent of the dimensionality for
$d \leq 2$, or is it just that the one and two-dimensional
exponents incidentally take very similar numerical values?
 The likely possibility of them being equal is quite intriguing
 since the return probabilities in systems with absorbing
 states are expected to be rather different in different 
spatial dimensions. 
 For the contact process, in dimensions larger 
than $d=2$ we find, as expected, dimension dependent exponents.
 
The fact that 
$\theta_{g}^{ZGB} > \theta_{l}^{ZGB}$ (in agreement
with field theoretical predictions \cite{oerd}
 is in marked contrast 
to well established findings in the field of reversible transitions, 
where quite generically $\theta_{g} < \theta_{l}$. A physical
explanation for such a discrepancy has been provided. 

We have also observed that power laws for persistence
are obtained only at criticality, and the associated 
exponents do not depend upon the considered initial 
conditions, {\it ad hoc} with what is known for other well 
known reversible systems.

 Persistence probabilities have also been 
evaluated at the first-order IPT of the ZGB model.
 Due to the short-lived behavior of 
the persistent sites we obtained an  exponent 
$\theta_{l}^ \simeq 3.0 \pm 0.3$. However, further 
effort will be required in order to clarify if this 
apparent power-law behavior merely reflects a short-time
regime followed by a cut-off or it is a real 
asymptotic effect.

We expect our numerical findings to constitute a valuable
step for the development of a theoretical framework in the
field of irreversible phase transitions, which is certainly
needed.
\vspace{0.3cm}

 {\bf\centerline{ACKNOWLEDGMENTS}}

 We acknowledge useful suggestions and discussions with
A. Gabrielli, P. Hurtado, and P. L. Garrido. 
 We thank also A. Baldasarri and S. Majumdar 
for very valuable comments and a critical reading of the manuscript.
E. A acknowledges the kind hospitality of Prof. J. Marro during
his stay at the University of Granada where the major
part of this work was performed. This work has been
financially supported by
CONICET, UNLP, CIC (As. As.), ANPCyT (Argentina) and the Volkswagen
Foundation (Germany), DGESIC (Spain) project PB97-0842, and
 the
European Network Contract ERBFMRXCT980183.

\pagebreak

\begin{figure}
\narrowtext
\centerline{{\epsfysize=3.0in \epsffile{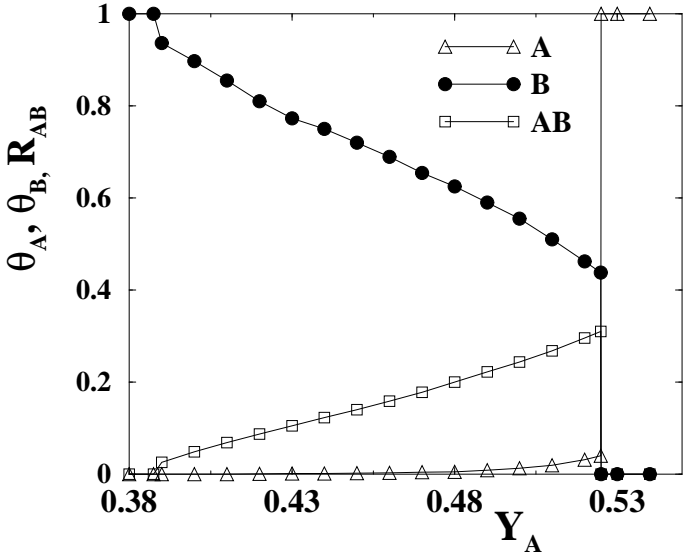}}}
\caption{Average coverage of $B-$species ($\theta_{B}$), $A-$species, 
($\theta_{A}$) and the $AB$ production rate ($R_{AB}$),
respectively, obtained under steady state operation, 
as a function of $Y_A$ for the ZGB model. 
Irreversible transitions of second- and first-order 
occurs at $Y_{1A} \simeq 0.387368$ and $Y_{2A} \simeq 0.52554$, 
respectively}.                
\label{fig1}
\end{figure}


\begin{figure}
\narrowtext
\centerline{{\epsfysize=3.0in \epsffile{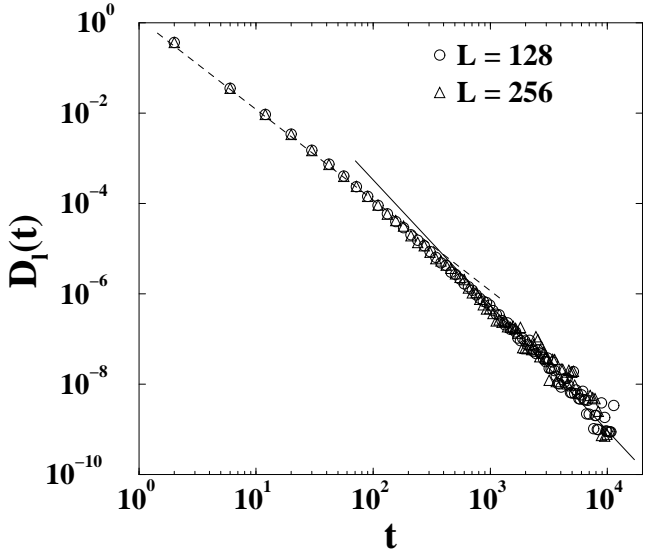}}}
\caption{ Log-log plot of $D_{l}(t)$ versus $t$, 
corresponding to the ZGB model at the second-order IPT
 and obtained for lattices 
of size  $\bigcirc, L = 128$
 and $ \bigtriangleup, L = 256$ , respectively.
The dashed (full) line shows the short- (long-)time regime  
and has slope $\theta_{l} +1 = 2.00$
($\theta_{l}^{ZGB} +1 = 2.5$ ), respectively. 
Results are obtained starting
 simulations with $\theta_{B}(t=0) = 0.1$ 
and averaging over $10^4$
 ($2.5 \times 10^3$) realizations for $L =  128$ 
($L= 256$), respectively.}   
\label{fig2}
\end{figure}

\begin{figure}
\narrowtext
\centerline{{\epsfysize=3.0in \epsffile{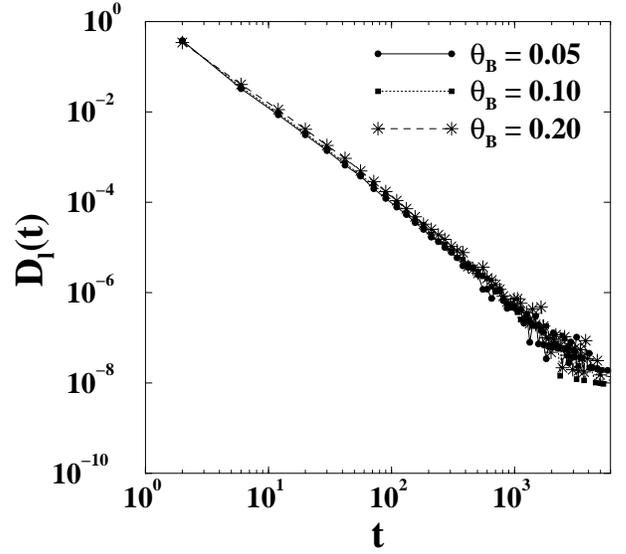}}}
\caption{ Log-log plot of $D_{l}(t)$ versus $t$, 
corresponding to the ZGB model at the second-order IPT
 and obtained for lattices 
of size $L = 256$.
Results obtained starting simulations with different values 
of $\theta_{B}(t=0)$, as indicated in the figure,  
and averaging over $2.5 \times 10^3$ realizations.
 No significative modification in the slope is observed 
upon changing the initial condition.}
\label{fig3}
\end{figure}

\begin{figure}
\narrowtext
\centerline{{\epsfysize=3.0in \epsffile{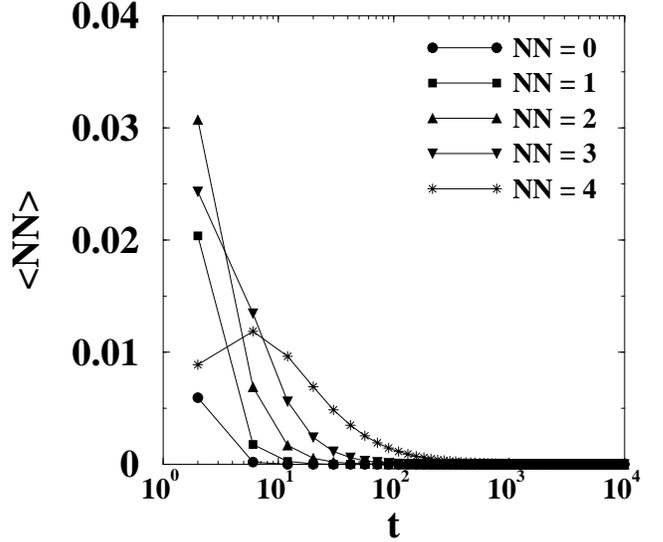}}}
\caption{ Linear-logarithmic plot of the average fraction of 
persistent sites with $NN$ empty neighbors, for $NN=0,1,2 3$, 
and $4$ respectively, as a function of time $t$.
Results are obtained using the same conditions as in figure 2.
 For times larger than $10^3$ the fraction of empty sites neighboring
persistent ones becomes very small; i.e. typically,
activity takes place away from persistent sites,
and local outbursts of activity are required in order
to reach persistent sites.} 
\label{fig4}
\end{figure}

\begin{figure}
\narrowtext
\centerline{{\epsfysize=3.0in \epsffile{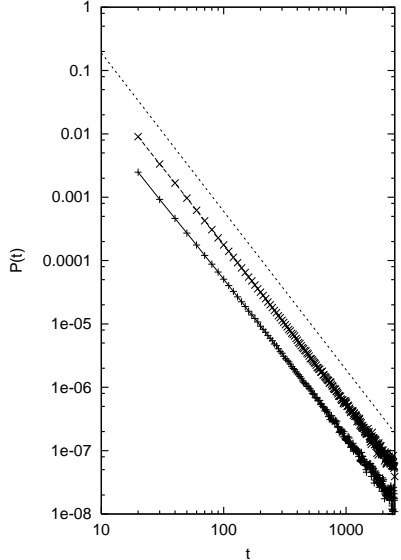}}}
\caption{
 Local persistence for one and two dimensional simulations
of the contact process. In $d=1$ ($d=2$) the system size
is $L=10^4$ ($256*256$) respectively. Observe that both
slopes are indistinguishable. The straight line has a slope 
$\theta_l=-1.5$.}
\label{fig5}
\end{figure}

\begin{figure}
\narrowtext
\centerline{{\epsfysize=3.0in \epsffile{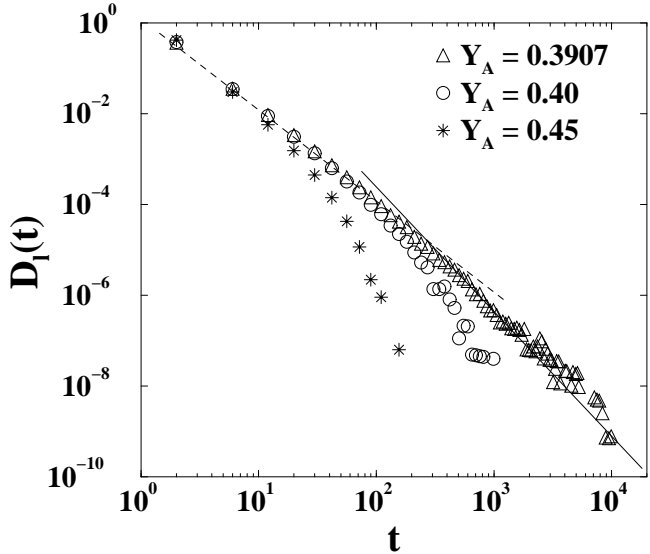}}}
\caption{ Log-log plot of $D_{l}(t)$ versus $t$, 
corresponding to the ZGB model. Results
obtained using lattices 
of size $L = 256$, with  $\theta_{B}(t=0) = 0.10 $,  
and averaging over $2.5 \times 10^3$ realizations.
Different values of the parameter $Y_{A}$ were used, as indicated in the 
figure. 
The dashed (full) line shows the short- (long-)time regime  
and has slope $2.00$
($\theta_{l}^{ZGB} +1 = 2.5$ ), and have been drawn 
to allow comparison with figure 2.
See more details in the text.}   
\label{fig6}
\end{figure}

\begin{figure}
\narrowtext
\centerline{{\epsfysize=3.0in \epsffile{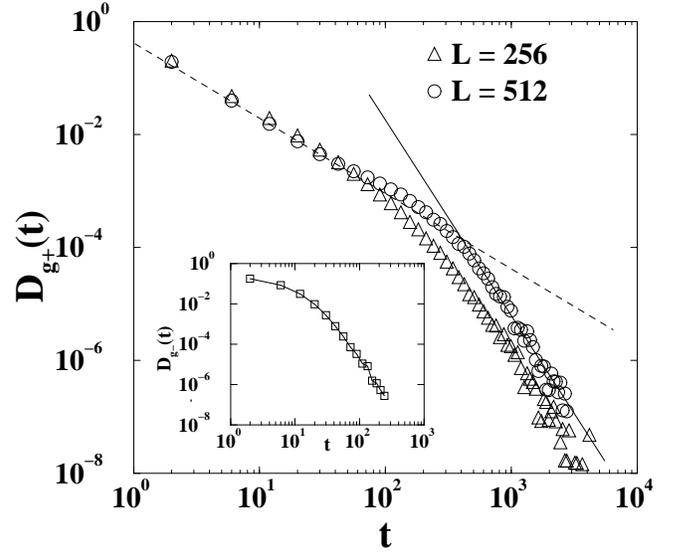}}}
\caption{Log-log plot of $D{g+}(t)$ versus $t$ 
corresponding to the ZGB model at the second-order IPT
 and obtained for 
lattices of side $\bigtriangleup, L = 256$ and $\bigcirc, L = 512$,  
respectively. The dashed (full) line
shows the short- (long-)time regime and has slope 
$1.25$, ($\theta_{g}^{ZGB} +1 = 3.5$), respectively.
Results have been obtained starting the simulations with 
$\theta_{A}(t=0) = 0.10$ and averaging  over $10^6$          
($2.5 \times 10^5$) realizations for $L = 256$ ($L = 512$), 
respectively.
The inset shows a log-log plot of $D_{g-}(t)$ versus $t$, obtained 
for $L = 512$: the asymptotic slope coincides with that of the
main plot.}
\label{fig7}
\end{figure}

\begin{figure}
\narrowtext
\centerline{{\epsfysize=3.0in \epsffile{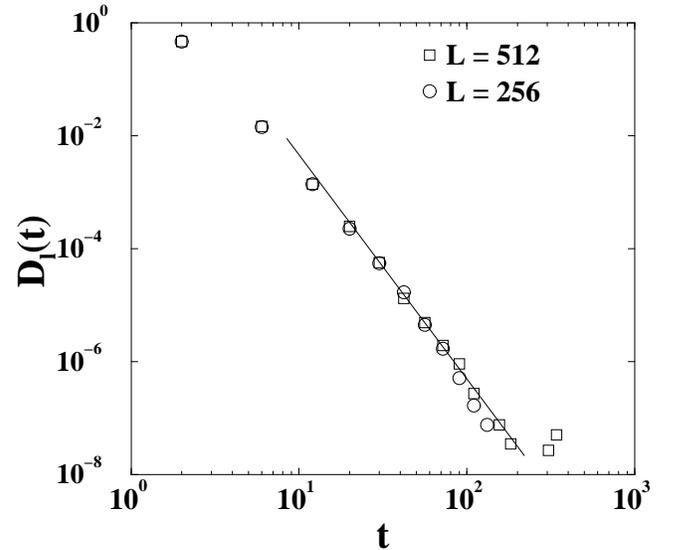}}}
\caption{ Log-log plot  
of $D_{l}(t)$ versus $t$ 
obtained at the coexistence point
($Y_{2A} = 0.52554$) for lattice sizes 
$L = 256$ and $L = 512$. 
The straight line has slope $\theta_{l} + 1 = 4.00$.
Results have been obtained starting simulations
 with $\theta_{A}(t=0) = 0.1$ 
and averaging over $10^4$ ($2.5 \times 10^3$) realizations for $L = 256$ 
($L= 512$), respectively. The upward bending of the last two points
indicate that the value of the critical point we are considering
is slightly in the subcritical phase for the finite system sizes
under analysis.}  
\label{fig8}
\end{figure}

\end{multicols}
\end{document}